\documentclass[%
 aip,
 amsmath,amssymb,
preprint,%
]{revtex4-1}

\usepackage{times}
\usepackage{amsmath}
\usepackage{amssymb}
\usepackage{amsfonts}
\usepackage{amsthm}
\usepackage{float}
\usepackage{bm}
\usepackage{latexsym}
\usepackage{hyperref}
\usepackage{multirow}
\usepackage[capitalize]{cleveref}
\usepackage{graphicx}
\usepackage{epstopdf}
\graphicspath{{./}} 
\hypersetup{
    colorlinks,
    citecolor=blue,    filecolor=blue,
    linkcolor=blue,    urlcolor=blue
}
\usepackage{lineno}

\begin{document}


\title{Transition signatures for electron-positron pair creation in space-time inhomogeneous electric field}


\author{C.~K.~Li}
\affiliation{Laboratory of Zhongyuan Light, School of Physics, Zhengzhou University, Zhengzhou 450001, China }
\author{X.~X.~Zhou}
\affiliation{School of Management Science and Engineering, Anhui University of Finance and Economics, \\
Bengbu 233030, China}
\author{Q.~Chen}
\affiliation{National Supercomputing Center in Zhengzhou, Zhengzhou University, Zhengzhou 450001, China}
\author{B.~An}
\affiliation{State Key Laboratory for Tunnel Engineering, China University of Mining and Technology, Beijing 100083, China}
\author{Y.~J.~Li}
\affiliation{State Key Laboratory for Tunnel Engineering, China University of Mining and Technology, Beijing 100083, China}
\affiliation{School of Science, China University of Mining and Technology, Beijing 100083, China}
\author{N.~S.~Lin}
\thanks{phy.nslin@gmail.com}
\affiliation{School of Science, China University of Mining and Technology, Beijing 100083, China}
\author{Y.~Wan}
\thanks{yangwan23@zzu.edu.cn}
\affiliation{Laboratory of Zhongyuan Light, School of Physics, Zhengzhou University, Zhengzhou 450001, China }


\date{\today}
\begin{abstract}
The process of electron-positron pair creation through multi-photon absorption in a space-time dependent electric field is analyzed using computational quantum field theory. Our findings reveal two distinct pair creation channels: the symmetric and asymmetric transition channels. We propose that the asymmetric transition channel arises from the inherent spatial inhomogeneity of intense laser pulses. By mapping the field-theoretical model of laser-assisted multi-photon pair creation onto a quantum-mechanical time-dependent framework, a semi-analytical solution that captures the asymmetric transition signatures of vacuum decay is derived. Additionally, it is demonstrated that neglecting spatial inhomogeneity leads to erroneous transition amplitudes and incorrect identification of pair creation channels. Furthermore, we have established that asymmetric transition channels substantially enhance the creation of electron-positron pairs for a given laser pulse energy.
\end{abstract}


\maketitle

\section{Introduction}
The possibility of creating electron-positron pairs from the quantum vacuum state through the interaction with a supercritical external field is one of the most striking predictions of quantum electrodynamics\cite{ADP1, AFed, YTP}.
There are two intrinsically different mechanisms by which electron-positron pairs can be created from the vacuum. 
The first scheme requires the field to be extremely strong and can be visualized as a quantum tunneling process between energy-shifted Dirac states, known as the Schwinger mechanism\cite{Schwinger}. 
The second scheme requires a very large laser frequency to trigger transitions between positive- and negative-energy states, referred to as the multi-photon mechanism\cite{Brezin}.
Schwinger pair creation has not been directly verified in experiments due to the current highest laser intensity\cite{JIN1}, approximately $10^{23}~\text{W/cm}^2$, is still significantly lower than the critical laser intensity of about $10^{29}~\text{W/cm}^2$.
If the energy of multiple laser photons, $n\omega$ , exceeds the gap between positive- and negative-energy continua, $2m_ec^2=1.022MeV$, then multi-photon pair creation can occur.
The perturbative multi-photon pair production was observed at the Stanford Linear Accelerator Center (SLAC) E144 using nonlinear Compton scattering\cite{Burke}.
However, in this laser-electron collision, the findings were not conclusive, as the electron-positron pairs could be  produced through either the Bethe-Heitler pair creation process or the multi-photon Breit-Wheeler process.
Following this pioneering experiment, several laboratories around the world, including the Extreme Light Infrastructure\cite{ELI}, the Center for Relativistic Laser Science\cite{CRLS}, the SLAC\cite{SLAC}, the Rutherford Appleton Laboratory\cite{RAL}, the European X-Ray Free-Electron Laser\cite{XRFEL} and Shanghai Superintense Ultrafast Laser Facility\cite{SULF}, are actively developing new methods to probe the quantum vacuum properties using very intense electromagnetic fields.

\begin{figure}[htbp]
\includegraphics[width=0.3\textwidth]{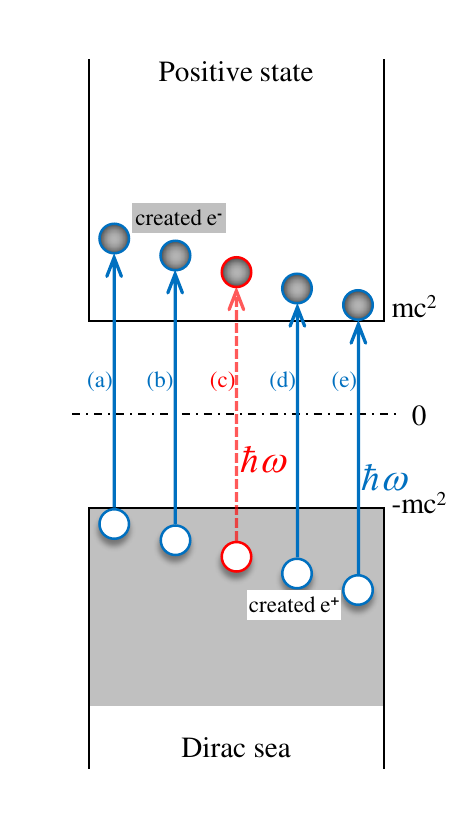}
\caption{\label{fig1}The scheme of multi-photon electron-positron pair creation. The asymmetrical transition are represented by (a), (b), (d), and (e), while the symmetrical transition is shown in (c).}
\end{figure}

In addition to experimental efforts, researchers have explored electron-positron pair creation using theoretical methods. The multi-photon pair creation process has been extensively studied by modeling the laser as a spatially homogeneous time-dependent electric field\cite{SSB1,CK1,HT1,RZJ1}. 
However, realistic ultra-strong fields, generated by short-pulse, focused lasers, introduce significant spatial inhomogeneity, which plays a critical role in pair creation. 
Several studies have examined the intriguing phenomena of pair creation induced by spatiotemporally inhomogeneous fields\cite{MF1, KK1, AW1, CK2, QZL1, CK3, QC1, LNH1, MJ1}. 
One of the key observations is the occurrence of asymmetric multi-photon transitions, which differ from the more commonly observed symmetric transitions. 
In intuitive terms, symmetric transition involves an electron moving from an initial negative-energy state $E_n=-\hbar\omega/2$ to a final positive-energy state $E_p=\hbar\omega/2$ (see the red dashed line in Fig.~\ref{fig1}). 
Asymmetric transition, on the other hand, involves an electron transitioning from $E_n \neq -\hbar\omega/2$ to $E_p \neq \hbar\omega/2$ with the satisfaction of energy conservation $E_p - E_n = \hbar\omega$ (see the blue solid lines in Fig.~\ref{fig1}).

Recently, asymmetric transitions were observed in an idealized external field composed of a temporally periodic vector field $A(t) = A_0 \cos(\omega t)$ and a spatially periodic scalar field $V(z) = V_0 \cos(kz)$\cite{MJ1}. 
However, such a field structure is highly specialized and challenging to realize with practical laser pulses. 
In this paper, we present a generalized model demonstrating that the asymmetric transition channel is directly related to the spatial width of a single external field.
The creation process under spatiotemporal variable field can also be described by the computational quantum field theory, which solves the Dirac equation numerically with full time and space resolution and allows us to compare and accurately gauge the quantum time-dependent perturbation theory.
In our study, we introduce the transition probability for electrons in the negative states and find noticeable asymmetric transition characteristics.
These characteristics are explained by relating the quantum field approach to the quantum mechanical perturbation description.
Moreover, we study the electron-positron pair creation from different spatial profile laser pulse which is have same energy.

This paper is organized as follows. 
In Sec.~\ref{2} we review the numerical and analytical approach to calculate the created electron-positron pairs from vacuum and present the external electric fields discussed in this paper.
In Sec.~\ref{3}, we present our detailed discussions on the asymmetric multi-photon transition phenomenon basen on the analytical and numerical result as well as optimal the pair creation induced by a laser focus field.
A summary is given in Sec.~\ref{4}.

\section{Theoretical Framework and The External Field}
\label{2}
\subsection{The computational quantum field theory}
Let us first briefly summarize our approach based on computational quantum field theory\cite{PK1,TC1,CKL1}. 
The interaction of the electric field with the electron-positron quantum field operator is described by the usual Dirac Hamiltonian
\begin{equation}
H=c{\sigma}_{1}[p_{x}-qA(x,t)/c]+ mc^{2}{\sigma}_{3},
\label{Eq1}
\end{equation}
where $\sigma_{1}$ and $ \sigma_{3}$ are the $2 \times 2$ Pauli matrices, $m$ and $q$ are the electron's mass and charge, and $c$ is the speed of light.
The characteristic spatial and temporal scales for the dynamics are naturally provided by the electrons’ Compton wavelength $\lambda_c \equiv \hbar/(mc)=3.9 \times 10^{-13} m$ and the corresponding time $T \equiv \hbar/(mc^2)=1.3 \times 10^{-21} s$. 
The field operator can be expanded as 
\begin{equation}
\begin{aligned}
\hat{\psi}(x, t)&=\sum_{p} \hat{b}_{p}(t)u_{p}(x)+\sum_{n} \hat{d}_{n}^{\dagger}(t)v_{n}(x)\\
&=\sum_{p}\hat{b}_{p}u_{p}(x,t)+\sum_{n} \hat{d}_{n}^{\dagger}v_{n}(x,t).
\label{Eq2}
\end{aligned}
\end{equation}
Here, $\hat{b}_{p}(t)$, $ \hat{d}^{\dagger}_{n}(t)$, $\hat{b}_{p}$ and $ \hat{d}^{\dagger}_{n} $ are the particle annihilation and the anti-particle creation operators, respectively.
The field-free Hamiltonian $H_{0}=c\sigma_{1}p_{x}+m\sigma_{3}c^{2}$ at $ t=0$, and its energy eigenstates are $ u_{p}(x) $ ($E \geq m c^{2}$) and $v_{n}(x)$ ($E \leq-m c^{2}$). 
$u_{p}(x,t)$ and  $v_{n}(x,t)$ are respectively the time evolved states of $u_{p}(x)$ and $ v_{n}(x)$, and their evolution satisfy the Dirac equation $i\hbar \partial \hat{\Psi}(x,t)= H \hat{\Psi}(x,t)$. 
The time evolution of the fermion annihilation and creation operators are 
\begin{equation}
\begin{aligned}
\hat{b}_{p}(t)=\sum_{p^{'}} \hat{b}_{p^{'}}\langle p|U(t)|p'\rangle+\sum_{n^{'}} \hat{d}^{\dagger}_{n^{'}}\langle p|U(t)|n'\rangle,\\
\hat{d}^{\dagger}_{n}(t)=\sum_{p^{'}} \hat{b}_{p^{'}}\langle n|U(t)|p'\rangle+\sum_{n^{'}} \hat{d}^{\dagger}_{n^{'}} \langle n|U(t)|n'\rangle.
\label{Eq3}
\end{aligned}
\end{equation}
Here, the anti-commutation relations $ [\hat{b}_{p},\hat{b}^{\dagger}_{p^{'}}]_{+}=\delta_{p,p^{'}},\quad[\hat{d}_{n},\hat{d}^{\dagger}_{n^{'}}]_{+}=\delta_{n,n^{'}}$ are applied.
$U(t)$ represents the time evolution operator that governs the evolution of the quantum state over time according to the Dirac equation, $|p\rangle=u_{p}(x)$ and $|n\rangle=v_{n}(x)$.
The time dependence of the total number of created electron-positron pairs is defined by the vacuum expectation value of products of the creation and annihilation operators 
\begin{equation}
N_{tot}(t)=\sum_{p,n}\langle\langle\mathrm{vac}\|\hat{b}^{\dagger}_{p}(t)\hat{b}_{p}(t)\|\mathrm{vac}\rangle\rangle=\sum_{p, n}\left|U_{p,n}(t)\right|^{2}.
\label{Eq4}
\end{equation}
These solutions of the space-time dependent Dirac equation with the vector potential $A(x,t)$ can be obtained on a space-time lattice using efficient fast-Fourier transformation based split-operator techniques.

In computational quantum field theory, an intuitive graphical interpretation of the electron-positron pair creation process is possible. 
The transition amplitude is determined by projecting the negative energy states $|E_n(t)\rangle$ onto the initial positive energy states $|E_p\rangle$ at time $t$. 
In this scenario, all external fields are simultaneously turned off at a specific time $t$, and the resulting time-dependent number of created particles represents the actual physical count of particles\cite{CCG1}.
In our method, $\left|U_{p,n}(t)\right|^{2}$ represents the expectation value of the number of created pairs with momentum $p$ and $n$. 
The energy of the positive state is given by $E_p=\sqrt{p^2c^2+c^4}$, while the energy of the negative state is $E_n=-\sqrt{n^2c^2+c^4}$.
Therefore, $\left| \langle E_p|E_n(t)\rangle \right|^{2}=\left|U_{E_p,E_n}(t)\right|^{2}$ can be interpreted as the probability of an electron transitioning from a negative energy state $|E_n\rangle$ to a positive energy state $|E_p\rangle$ at time $t$.

\subsection{The external electric field}
In this work, we utilize a spatiotemporally inhomogeneous electric field characterized by a peak field strength $A_0$, space width $\lambda$, pulse duration $\tau$, and oscillation frequence $\omega$ of the form
\begin{equation}                  
A(x,t)=A_0exp(-x^2/2\lambda^2)exp(-t^2/2\tau^2)cos(\omega t).
\label{Eq5}
\end{equation}
Here, the $x$-direction denotes the direction in which the amplitude of the electric field varies.
The field can be viewed as a model of the electric field in a standing-wave pattern produced by two counter-propagating laser pulses. 
In this setup, the electric field is generated at the anti-nodes of the pattern, where the associated magnetic field vanishes, representing the finite focus of a laser pulse \cite{CK2}. 
Notably, in this configuration, there are no propagating waves present.

\subsection{The time-dependent perturbation theory}
\label{2C}
We consider first-order time-dependent perturbation theory to theoretically describe the single-photon transition. 
Since the interaction does not couple spin directions, we can express the field operator using only two components.
The Hamiltonian of this system can then be separated into two parts: $H=H_0+H^{'}$, where the external field is represented by $H^{'}=-q{\sigma}_{1}A(x,t)$, treated as a perturbation, and $H_0$ is the field-free Hamiltonian, analogous to the computational quantum field theory.
The first-order transition amplitude from the negative state $|n\rangle$ to the positive state $|p\rangle$ at an arbitrary time $t$ is given by 
\begin{equation}
C_{p,n}(t)=\int_{0}^{t}\langle p|H^{'}|n \rangle exp[i(E_p-E_n)\tau]/i~d\tau.
\label{Eq6}
\end{equation}
By summing over all the states of $p$ and $n$, we derive the first-order perturbation estimate of the total pair creation as 
\begin{widetext}
\begin{eqnarray}
N_{tot}^{'}(t)=\sum_{p, n}\left|C_{p,n}(t)\right|^{2}
=\sum_{p, n}|\kappa(\lambda) 
i\sqrt{2\pi}/2exp(-\tau^2\omega_{pn}/2)  
[erfi(\sqrt{2}\tau\omega_{pn}/2) 
-erfi(\sqrt{2}\tau\omega_{pn}/2+i\sqrt{2}t/2\tau)]|^2,
\label{Eq7}
\end{eqnarray}
\end{widetext}
where $\omega_{pn}=E_p-E_n-\hbar \omega$ denotes the energy conservation condition and $\kappa(\lambda)$ represents the coupling strength of the quantum states.
We derived the coupling strength of the quantum states as
\begin{equation}
\begin{aligned}
\kappa(\lambda)&=\langle p|-q{\sigma}_{1}A(x)|n \rangle \\
&=2\pi A_0^2\lambda^2A_{p,n}exp[-(p+n)^2\lambda^2/2]^2,
\label{Eq8}
\end{aligned}
\end{equation} 
where the inner product for spins $A_{p,n}$ is defined as $[sgn(n)\sqrt{E_p+c^2}\sqrt{-E_n-c^2}+sgn(p)\sqrt{E_p-c^2}\sqrt{-E_n+c^2}]/[4\pi\sqrt{-E_pE_n}]$.

\section{Results and Discussion}
\label{3}
\subsection{The asymmetric multi-phton transition}
\label{3A}
\begin{figure*}[htbp]
\includegraphics[width=0.45\textwidth]{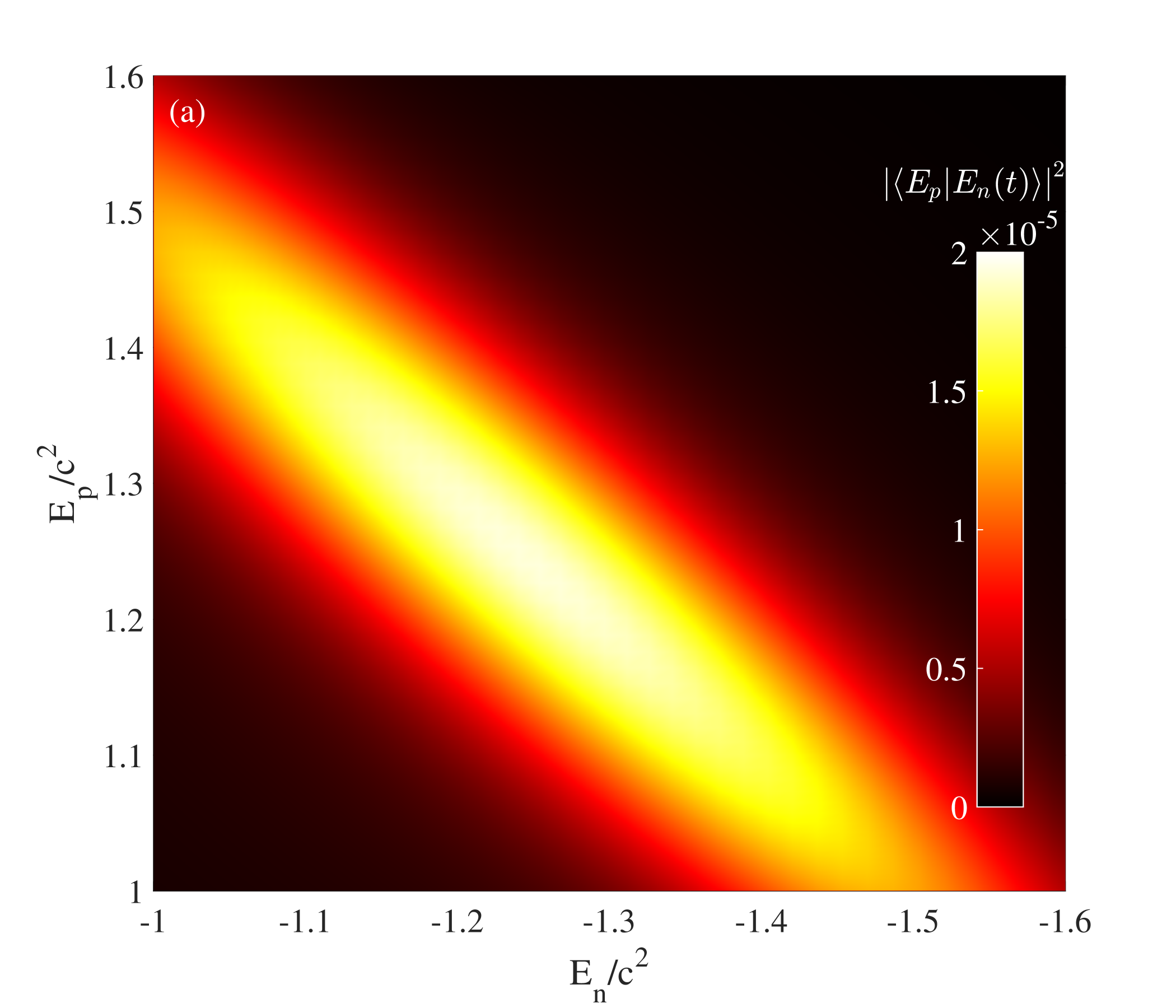}
\includegraphics[width=0.45\textwidth]{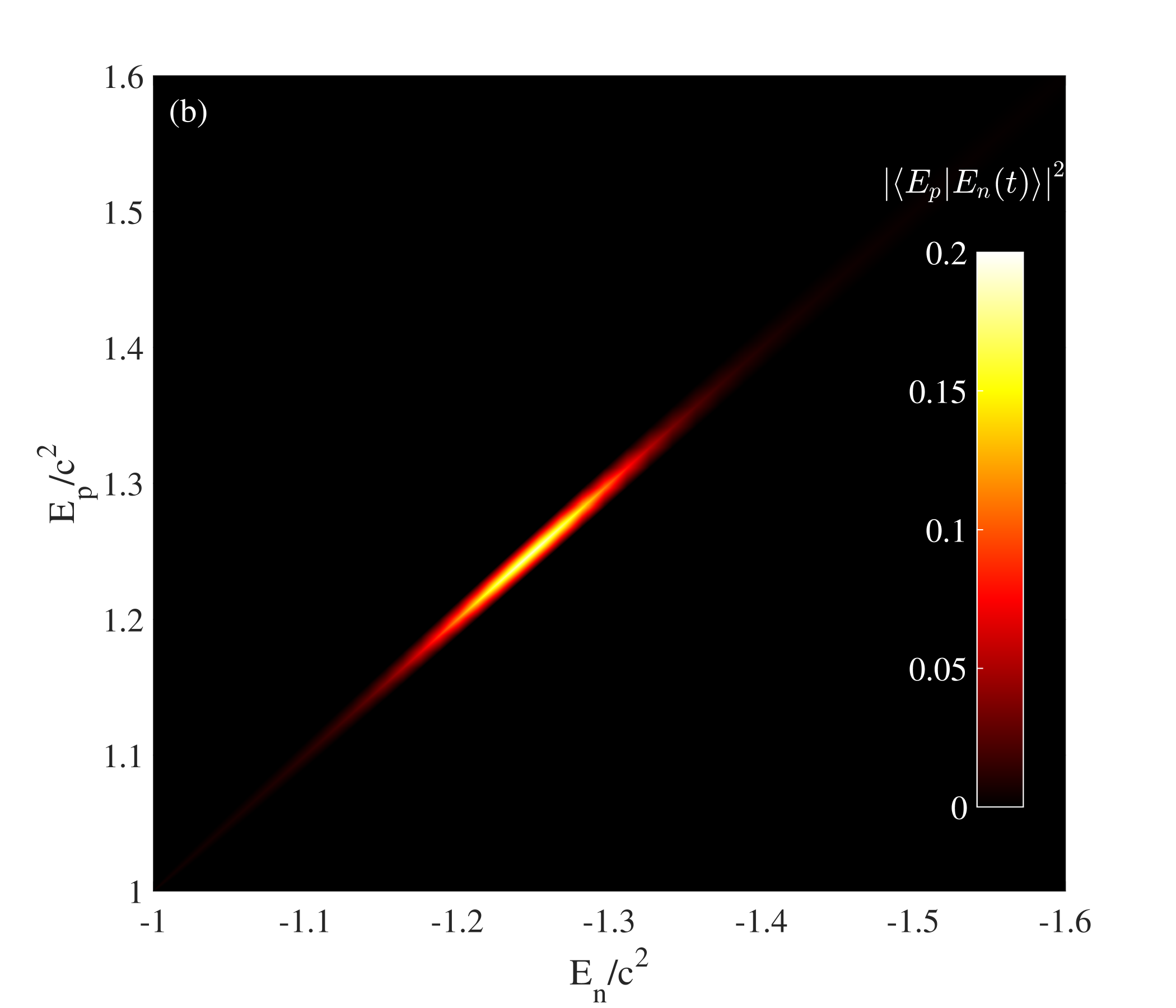}
\caption{\label{fig2}The transition probability of electron for different space width (a) $\lambda=\lambda_c$, (b) $\infty$. The parameters are $A_0=0.05mc^2$, $\hbar\omega=2.5mc^2$, $\tau=5 T_L=10 \pi/\omega$, $t=20 T_L=40 \pi/\omega$.}
\end{figure*}

The most remarkable effect is that the electron-positron pair creation process exhibits asymmetric transition characteristics, with an asymmetric transition scale determined by the spatial extent of the laser pulse, as shown in Fig.~\ref{fig2}.
The horizontal axis represents negative energy, while the vertical axis represents positive energy. 
The electron's transition probabilities for specific states are indicated by the color scale.
The “anti-diagonal line” in the southwest-to-northeast facing diagonal denotes symmetric transitions with opposite momentum (identical absolute value) and symmetric energy ($E_n=-E_p$). 
Conversely, the area outside this line denotes asymmetric transitions with different momentum and asymmetric energy ($E_n \neq E_p$).
Note that the transition probability of the electron exhibits a "block" shape, rather than being represented as a line (parallel to the diagonal line) dictated by the energy conservation condition $\omega_{pn}=E_p-E_n-\hbar\omega=0$.
Therefore, the multi-photon pair creation proceeds via both resonance and non-resonance transition paths \cite{MF1,GRM1}.
This phenomenon can be explained by deriving the first-order perturbation theory.
Based on the time-dependent perturbation theory, the energy conservation term in the total creation number $N_{tot}^{'}(t)$ (see Eq.~\ref{Eq7}) is a complex function $i\sqrt{2\pi}/2exp(-\tau^2\omega_{pn}/2)[erfi(\sqrt{2}\tau\omega_{pn}/2)-erfi(\sqrt{2}\tau\omega_{pn}/2+i\sqrt{2}t/2\tau)]$, rather than a delta function $\delta\omega_{pn}$.
The energy conservation term in Eq.~\ref{Eq7} leads to the transition probability of the electron taking on a "block" shape, which incorporates the non-resonant transition pathway.

In order to get a better idea how this asymmetric transition is possible, we performed simulations for two critical case.
In Fig.~\ref{fig2} (a), we illustrate the transition probability of the electron when the spatial width $\lambda$ is equal to the electrons’ Compton wavelength $\lambda_c$.
The electrons’ Compton wavelength $\lambda_c$ is the minimum spatial length for electron-positron pair creation.
In this case, electron-positron pairs are created by both the symmetric transition channels and the asymmetric transition channels, with the asymmetric transition channels accounting for the vast majority of the total yield.
For the case of maximal spatial width, $\lambda$ is infinite, as shown in Fig.~\ref{fig2} (b).
In this case ($\lambda=\infty$), the laser spatial width $\lambda$ is much larger than the Compton wavelength $\lambda_c$, which is the typical length scale of the electron-positron pair creation\cite{BSX1}. 
Therefore, the spatial extent of the laser pulse is neglected; namely, the laser pulse $A(x,t)$ is approximate to a spatially uniform and time-dependent field $A(t)$.
In our simulation, we use the electric field $A(t)=A_0exp(-t^2/2\tau^2)cos(\omega t)$ instead of the field $A(x,t)=A_0exp(-x^2/2\lambda^2)exp(-t^2/2\tau^2)cos(\omega t)$ for the case of infinite spatial width.
The space-independent electric field $A(t)$ vanishes the asymmetric transition channels and enhances the symmetric transition channels, as shown in Fig.~\ref{fig2} (b).
In other words, the asymmetric transition channels associated with $A(x,t)$ are absent in the data for $A(t)$.
It is suggested that the spatial inhomogeneity of the field can trigger numerous irreversible transitions from each negative energy state to each positive energy state, thereby opening the asymmetric transition channels.
Notably, it is a significant assumption that spatiotemporally inhomogeneous laser fields $A(x,t)$ can be approximated as a spatially uniform and time-dependent electric field $A(t)=A_0exp(-t^2/2\tau^2)cos(\omega t)$ when the laser spatial width $\lambda$ is much larger than the typical length scale of this process $\lambda_c$.
However, this assumption of spatial homogeneity leads to incorrect predictions regarding the creation of electron-positron pairs, as it fails to account for the asymmetric transitions that would occur.
In addition, the symmetric transition probability for this case is approximately four orders of magnitude greater than that when the spatial width $\lambda$ equals the Compton wavelength $\lambda_c$. 
This is because the laser energy throughout the entire numerical box is maximal when the spatial width is infinite.

\begin{figure}[htbp]
\includegraphics[width=0.45\textwidth]{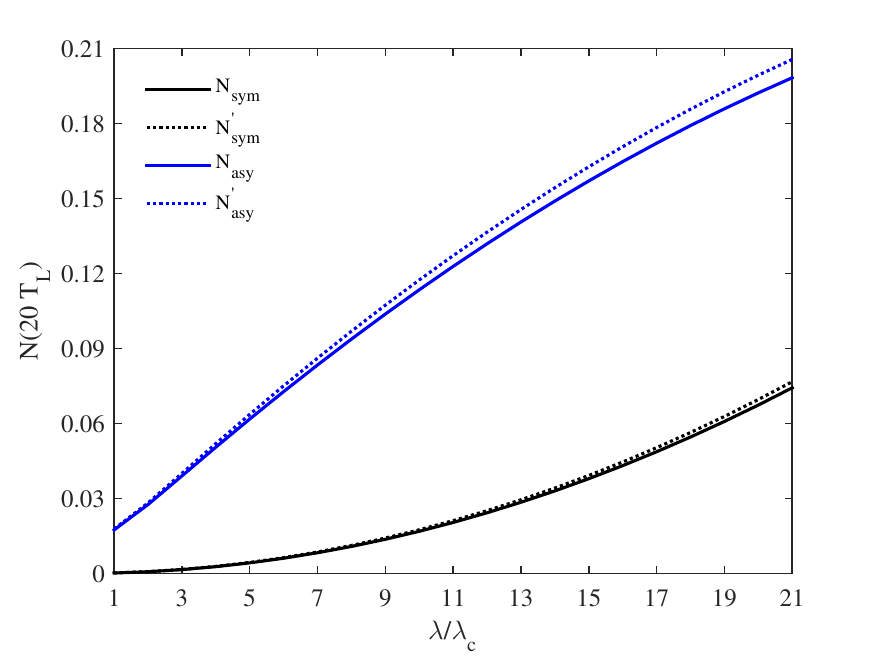}
\caption{\label{fig3} The number of the created pairs  as a function of field space extent $\lambda$  (in units of the electrons’ Compton wavelength $\lambda_c$). The black line denotes the number of pairs created by the symmetric channels, while the blue line represents the number of pairs created by the asymmetric channels. The solid line and the dashed line represent the results of computational quantum field theory and time-dependent perturbation theory, respectively. The parameters are $A_0=0.05mc^2$, $\hbar\omega=2.5mc^2$, $\tau=5 T_L=10 \pi/\omega$, $t=20 T_L=40 \pi/\omega$.}
\end{figure}

We have computed the pair-creation number $N_{sym}(t)$ and $N_{asy}(t)$ to examine the asymmetric transition process more quantitatively.
The term $N_{sym}(t)$ denotes the pair-creation number from the symmetric transition process, defined as $N_{sym}(t)=\sum_{p, n}\left|U_{p,n}(t)\right|^{2}, \left| p \right| = \left| n \right|$.
The number $N_{asy}(t)$ is defined as $N_{asy}(t)=\sum_{p, n}\left|U_{p,n}(t)\right|^{2}, \left| p \right| \neq \left| n \right|$, representing the electron-positron pairs created by the asymmetric transition paths.
In Fig.~\ref{fig3}, we depict the spatial dependence of the pair-creation numbers $N_{sym}(t=20T_L)$ and $N_{asy}(t=20T_L)$.
We see that, in a spatiotemporal inhomogeneous field, the electron-positron pairs are created through two different paths: the symmetric transition channels and the asymmetric transition channels.
As predicted in Fig.~\ref{fig2} (a), the asymmetric transition channels play a dominant role in the pair creation process when the spatial width $\lambda=\lambda_c$.
The number $N_{asy}(t=20T_L)$ is equal to 0.017 and accounting for $98.86\%$ of total number in this case. 
Furthermore, we have observed that increasing the spatial width $\lambda$ leads to a significant reduction in the slope of the function $N_{asy}(\lambda, 20T_L)$.
Conversely, increasing $\lambda$ enhances the slope of the function $N_{sym}(\lambda, 20T_L)$.
Based on Fig.\ref{fig2} and Fig.\ref{fig3}, we predict that the pair-creation number $N_{sym}(\lambda, 20T_L)$ equals $N_{asy}(\lambda, 20T_L)$ at a certain spatial width $\lambda$. 
Beyond this spatial width, $N_{sym}(\lambda, 20T_L)$ is anticipated to exceed $N_{asy}(\lambda, 20T_L)$ in magnitude.

To understand the physical origin of the asymmetric transition channels, we recall that, for a space-time dependent electric field, the pair creation process can be viewed as a time-dependent perturbation problem, as shown in Sec.~\ref{2C}. 
The perturbative estimates of the number of symmetrically created pairs $N_{sym}^{'}(t)$ and asymmetrically created pairs $N_{asy}^{'}(t)$ are obtained by summing over the corresponding states of $p$ and $n$.
The Fig.~\ref{fig3} also displays the space-dependent pair-creation number predicted by perturbation theory.
Similar to the results of computational quantum field theory, the pairs creation number $N_{asy}^{'}(\lambda, 20T_L)$ for the asymmetric transition process show slow growth with increasing spatial width $\lambda$, while $N_{sym}^{'}(\lambda, 20T_L)$ exhibits rapid growth.
Due to the fact that the field strength increases across the entire numerical box as $\lambda$ increases, the error of the computational quantum field theory and the first-order perturbation theory increases as the spatial width $\lambda$ increases.
However, the maximal error is $3.26\%$ and $3.7\%$  for the symmetric transition channels and the asymmetric transition channels, resprctively.
The excellent match of  $N_{sym}^{'}(\lambda, 20T_L)$ and $N_{asy}^{'}(\lambda, 20T_L)$ with the numerical data suggests that the first-order time-dependent perturbation theory can indeed capture the main features of the single-photon transition process.

\begin{figure*}[htbp]
\includegraphics[width=0.45\textwidth]{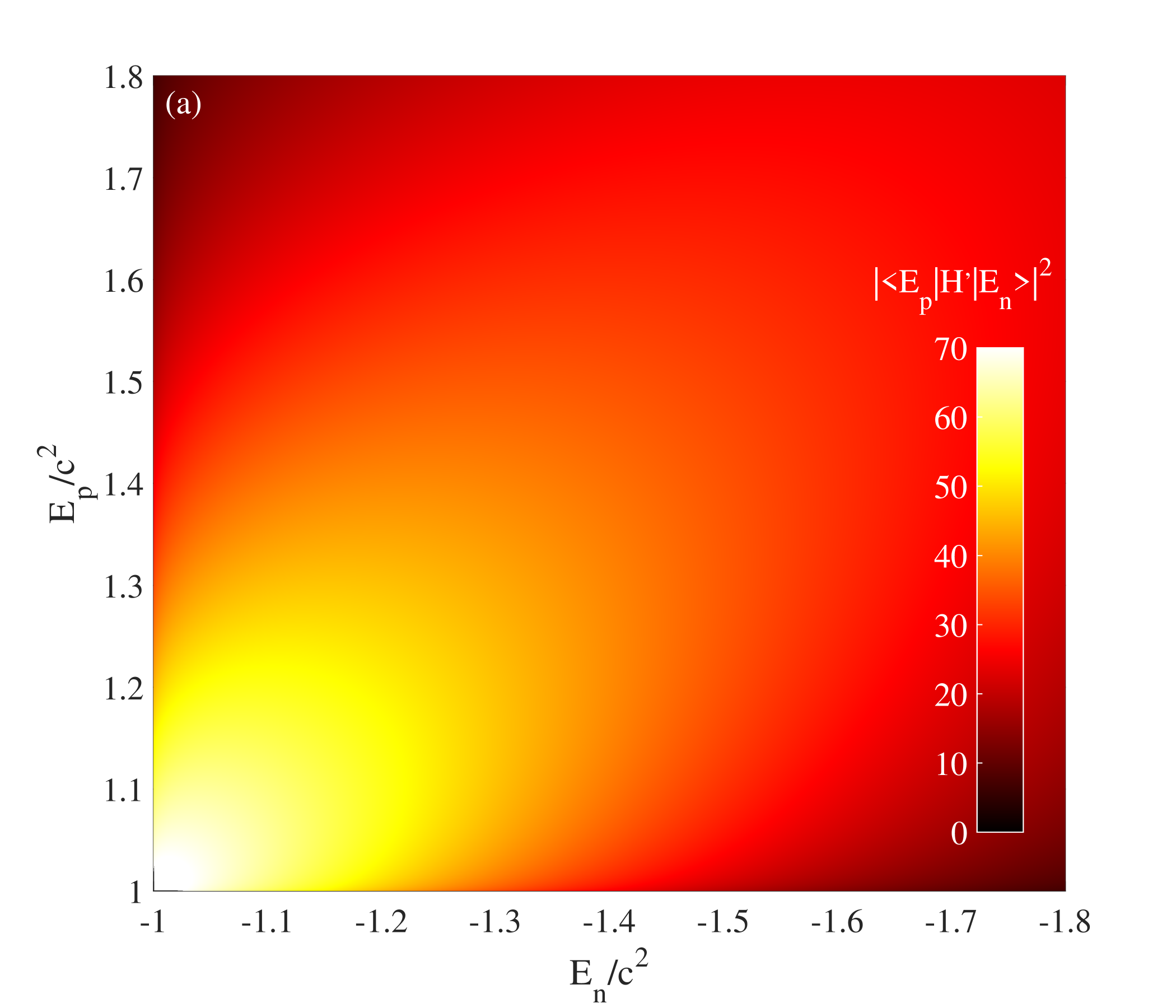}
\includegraphics[width=0.45\textwidth]{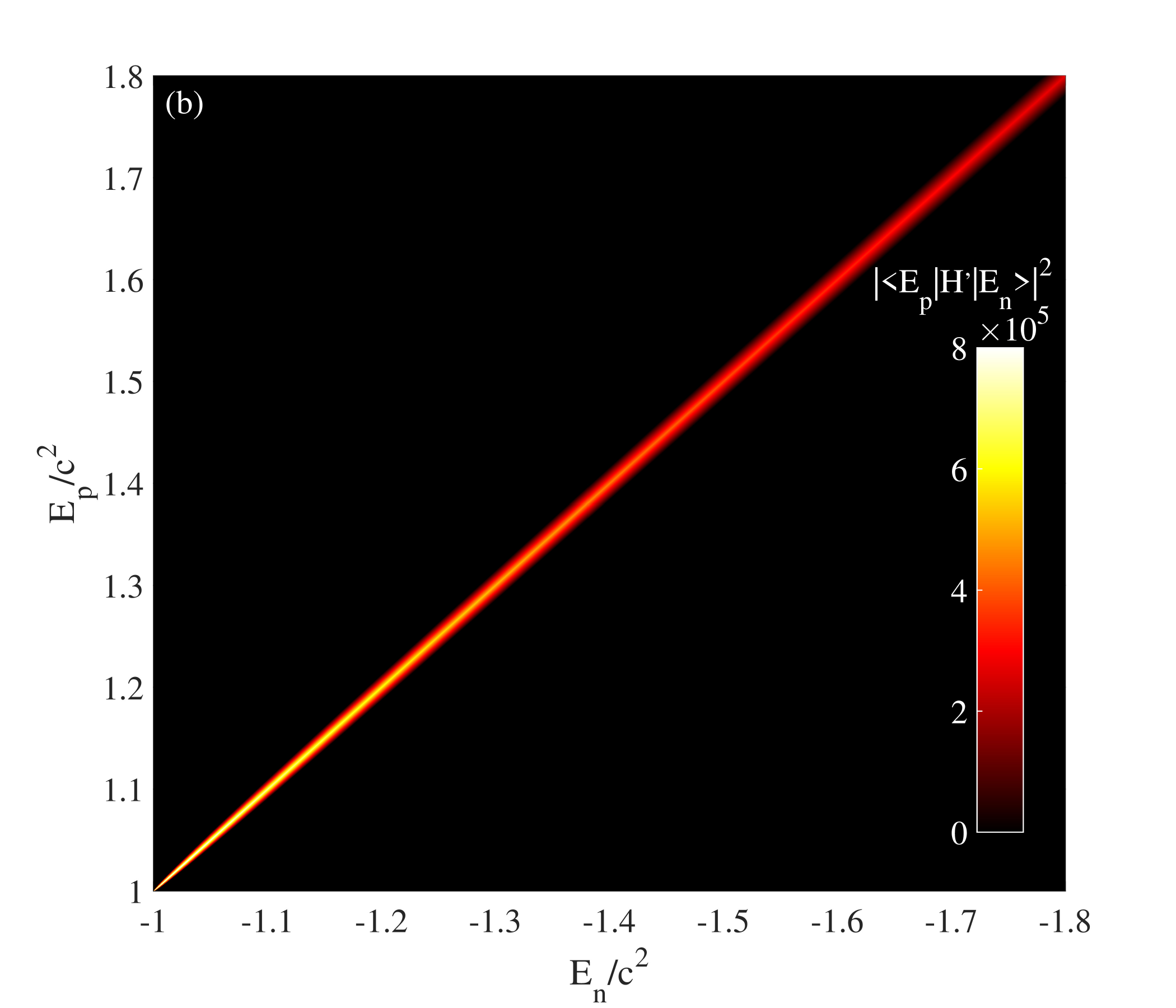}
\caption{\label{fig4}The coupling strengths $\kappa(\lambda,E_n,E_p)$ of initial energy state for different spatial width (a) $\lambda=\lambda_c$, (b) $\infty$. The parameters are $A_0=0.05mc^2$.}
\end{figure*}

This indicates that the asymmetric transition channels can be attributed to the spatial dependence of the coupling strengths $\kappa(\lambda)$ (see Eq.~\ref{Eq8}).
In Fig.~\ref{fig4}, we have graphed the coupling strengths $\kappa(\lambda,E_n,E_p)$ between the negetive energy state $E_n$ and the positive energy state $E_p$ for $\lambda=\lambda_c$ and $\lambda=\infty$. 
We observe that the coupling strength $\kappa(\lambda=\lambda_c,E_n,E_p)$ is universally applicable across the entire energy range, whereas the coupling strengths $\kappa(\lambda=\infty,E_n,E_p)$ is valid only for symmetric negative and positive states.
Comparing with Fig.~\ref{fig2}, asymmetric transitions are permitted for the space-dependent field $A(x,t)$, whereas asymmetric transitions are not permitted for the space-independent field $A(t)$.
It is immediately clear how asymmetric transition signatures arise in the more realistic case of additional spatial pulse inhomogeneities.
From a mathematical point of view, this exponential term $exp[-(p+n)^2\lambda^2/2]^2$ (see Eq.~\ref{Eq8}) in the coupling strengths $\kappa(\lambda=\infty,E_n,E_p)$ equals "1" only when $p+n=0$ and $exp[-(p+n)^2\lambda^2/2]^2=0$ for $p+n \neq 0$.
This indicates conservation of the total canonical momentum of the system, which exclusively opens symmetric transition channels.
For a finite spatial width of the field, $\lambda \neq \infty$, the exponential term $exp[-(p+n)^2\lambda^2/2]^2$ is non-zero for asymmetric negative and positive states.
This suggests that under these conditions, asymmetric transitions between negative and positive states are possible.
In other words, the asymmetric transition due to the spatial inhomogeneities of field $A(x,t)$ (compared to $A(t)$) becomes important.

\subsection{The optimization of electron-positron pair creation}
\label{3B}
\begin{figure}[htbp]
\includegraphics[width=0.45\textwidth]{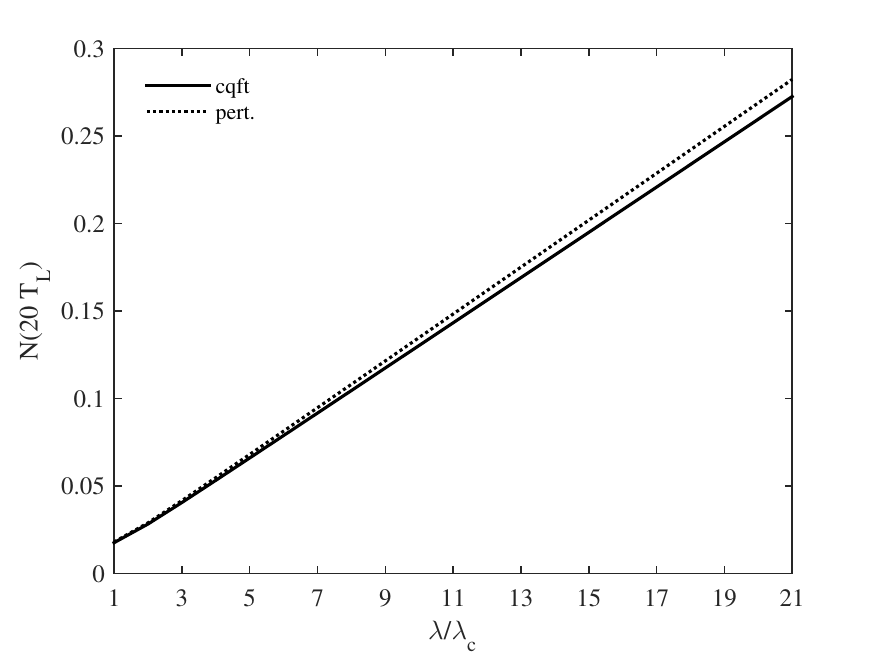}
\caption{\label{fig5} The total number of the created pairs as a function of field space extent $\lambda$ (in units of the electrons’ Compton wavelength $\lambda_c$). The solid line and the dashed line represent the results of computational quantum field theory and time-dependent perturbation theory, respectively. The parameters are $A_0=0.05mc^2$, $\hbar\omega=2.5mc^2$, $\tau=5 T_L=10 \pi/\omega$, $t=20 T_L=40 \pi/\omega$.}
\end{figure}

In this subsection, we examine the enhancement effect of asymmetric multi-photon transition channels on electron-positron pair production.
The dependence of the total number of created pairs on the field width is illustrated in Fig.~\ref{fig5}. 
Notably, the total number of pairs exhibits a linear increase with increasing field width.
This observed dependence on field width is counterintuitive and contrary to initial expectations.
Surprisingly, the model predicts that for smaller field width, where a greater number of transition channels are available, the particle number $N(20 T_L)$ is lower than for larger field width, where fewer transition channels are present.
This suggests that asymmetric multi-photon transitions suppress, rather than enhance, the creation of electron-positron pairs.
This finding stands in contrast to the results reported in Ref.~\cite{MJ1}, which indicate that asymmetric transition channels enhance pair creation.

\begin{figure}[htbp]
\includegraphics[width=0.45\textwidth]{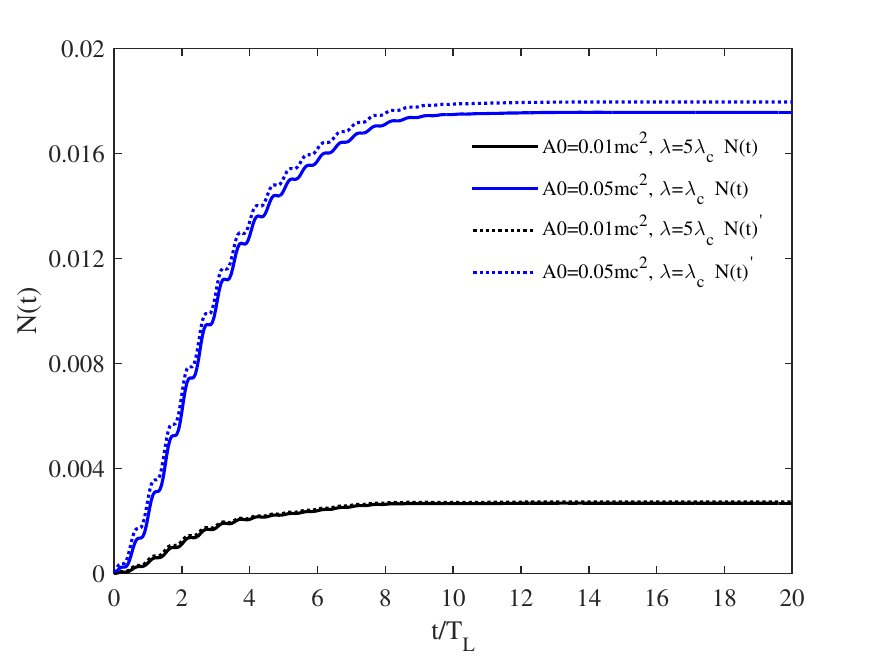}
\caption{\label{fig6} The number of the created pairs as a function of time. The black line denotes the laser parameters $A_0=0.05mc^2$, $\lambda=\lambda_c$, while the blue line represents the laser parameters $A_0=0.01mc^2$, $\lambda=5\lambda_c$. The solid line and the dashed line represent the results of computational quantum field theory and time-dependent perturbation theory, respectively. The other parameters are $\hbar\omega=2.5mc^2$, $\tau=5 T_L=10 \pi/\omega$.}
\end{figure}

To elucidate the impact of asymmetric multi-photon transitions on electron-positron pair creation, we conducted a comprehensive analysis of the laser energy distribution across the entire numerical domain.
We found that the total energy increases with the field space extent $\lambda$.
Given that the total energy of the external field is a critical factor in vacuum pair creation \cite{ADP1,AFed,YTP,Schwinger,Brezin}, we further investigated vacuum decay under conditions of constant laser energy.
In our model, the focal spot of the laser pulse is described by $A'(x)=A_0exp(-x^2/2\lambda^2)$.
The total energy of the pulse is defined as $H=\int_{+\infty}^{-\infty}A'(x)dx=\sqrt{2\pi}A_0\lambda$.
In order to ensure the total energy remained constant, we simulated two laser pulses with distinct spatial profiles: a narrow laser field with the peak field strength $A_0=0.05mc^2$ and the field width $\lambda=\lambda_c$, and a wide laser field with the peak field strength $A_0=0.01mc^2$ and the field width $\lambda=5\lambda_c$.
The Fig.~\ref{fig6} illustrates the time evolution of the number of created particles.
The irreversible growth occurs during interaction period, with the total number of created particles reaching $N(t=20T_L)=0.0176$ for the narrow laser field and $N(t=20T_L)=0.0027$ for the wide laser field.  
Notably, the narrow laser field creates nearly 6.5 times more particles than the wide field, despite the total energy being identical.
As discussed in Sec.~\ref{3A}, the spatial inhomogeneities of laser field $A(x,t)$ give rise to asymmetric multi-photon transition channels.
When the total energy is held constant, the narrow laser field exhibits a greater number of multi-photon transition channels compared to the wide field. 
Consequently, the narrow laser field generates a significantly larger number of electron-positron pairs than the wide field.
These results demonstrate that asymmetric multi-photon transition channels enhance electron-positron pair creation for a fixed total energy of the laser pulse.

\section{Conclusion}
\label{4}
In summary, electron-positron pair creation in spatially and temporally inhomogeneous electric fields has been studied both numerically and theoretically.
We have demonstrated that the spatial extent of the field suppresses the asymmetric transition pathways while enhancing the symmetric transition pathways.
From a theoretical viewpoint, we find that the spatial inhomogeneity of the field brings about the asymmetric multi-photon transition channels.
In addition, when the spatial width of a laser pulse is much larger than the characteristic length scale of the pair creation process, the laser pulse can be approximated as a spatially uniform time-dependent electric field.
However, the omission of the laser’s spatial inhomogeneity changes the entire pair-creation process and therefore leads to unreliable predictions.
Consequently, the transition amplitude of the electron is qualitatively incorrect and excludes the physical asymmetric transition process.
Furthermore, we observed that asymmetric transition pathways significantly increase the total yield of electron-positron pairs for the same laser pulse energy. 
Conversely, when the laser energy was varied, a reduction in the production of electron-positron pairs was observed, despite the narrower laser field offering more transition pathways.

\begin{acknowledgments}
Fruitful discussions with M. Jiang and W. M. Wang are gratefully acknowledged. This work has been supported by the National Natural Science Foundation of China (NSFC) under Grants $Nos. 12447120$ and $12204001$. Numerical simulations were implemented on the SongShan supercomputer at National Supercomputing Center in Zhengzhou.
\end{acknowledgments}


\end{document}